\documentclass[twocolumn]{jpsj2} 
\usepackage{bm}

\title{
          Vortex oscillations in 
          confined Bose-Einstein condensate interacting with 1D optical lattice
}

\author{Tomoya \textsc{Isoshima}}

\inst{Institute of Physics, Academia Sinica, Nankang, Taipei, Taiwan 11529}

\abst{
We study Bose-Einstein condensate of atomic Boson gases trapped in
a composite potential of a harmonic potential and an
optical lattice potential.
We found a series of collective excitations
that induces localized vortex oscillations with a characteristic wavelength.
The oscillations might be observed experimentally when radial confinement is tight.
We present the excitation spectra of the vortex oscillation modes
and propose a way to experimentally excite the modes.
}

\kword{vortex, optical lattice, Bose-Einstein condensate}

\begin{document}
\maketitle

\section{Introduction}

Since the realization of vortex state
in Bose-Einstein condensates (BEC) in traps~\cite{matthews,madison},
various kinds of vortex configurations 
are experimentally created.
A clear advantage of studying vortices 
in the condensates of atomic gases is that
the vortex cores are directly observable.
Using the imaging of atom density in planes perpendicular 
to the vortex axis, 
vortex lattices,~\cite{lattice:spinor} giant vortex,~\cite{giant}
core-less vortex,~\cite{coreless} and
doubly quantized vortices~\cite{doubly:1,doubly:2} have been observed. 
By the imaging parallel to the vortex lines the bending~\cite{perp} and
the oscillation~\cite{kelvon} of the vortex line are found.

Recently dilute atomic gases interacting with the
optical lattice~\cite{pitaevskii-stringari}
are actively studied. 
The Boson gases in the optical lattice behave in two different ways,
as a superfluid or a Mott-insulator.~\cite{mott}
The condition of the phase transition between these two phases is
determined by various parameters, such as the particle number and
the intensity and the dimension of the lattice potential.
Vortices are also an interesting phenomena here.
Vortices may exist
in presence of the optical lattice potentials,
and the properties of the vortices are affected by the lattice potential.
There also are researches on several kinds of the vortices 
on the BEC in the optical lattices 
near the Mott-insulator transition~\cite{congjun:wu}
and in the superfluid
region~\cite{vortices:optical:lattice,martikainen-stoof,martikainen-stoof2}.

The vortices whose cores are perpendicular to the 2D optical lattices
are studied in ref.~\citen{vortices:optical:lattice}.
The dynamics of vortices and a new kind of vortex that is unique 
to the optical lattices are reported.
On 1D optical lattice,
the vortices parallel to the optical lattice are
studied~\cite{martikainen-stoof,martikainen-stoof2} and
they predicted various vortex oscillations
utilizing the analytical formulations.
The formulation assumes periodic system along the lattice and
neglects the effects of the axial confinement potential.

In this paper, we treat the condensate in 1D optical lattice
confined along the principal axis,
to find features unique to the confined systems.
The whole confined system is treated
within the Gross-Pitaevskii (GP) and Bogoliubov equations,
and we study structures of the spectra that are unique to the confined system.

In our previous work, we found~\cite{iso-lattice} characteristic excitations
in the condensate confined in 1D optical lattice in absence of the vortex.
The branch of these characteristic excitations was identified
by a breaking down of phase count,
but the detailed origin and the properties were not clarified.
This paper studies corresponding modes, especially within the Kelvin modes
that accompanies vortices.
And
identify the characteristic excitations as the shortest wavelength modes
both in the vortex systems and the vortexfree systems.

\section{Formulation}

We study the excitation spectra of BEC
confined in a composite potential
\begin{align}
    V(r, z) &=  \frac{m}{2} (2\pi\nu_r)^2 r^2
       + V_z(z)
       + V_\mathrm{opt}(z),
\label{eq:v}
\\
    V_z(z) &=  \frac{m}{2} (2\pi\nu_z)^2 z^2, 
\label{eq:vz}
\\
    V_\mathrm{opt}(z) &=  s E_\mathrm{R} \mathrm{sin}^2 \left( \pi \frac{z}{d} \right)
\label{eq:vopt}
\end{align}
that consists of an optical lattice in the principal direction of consideration 
and harmonic confinement potentials along the lateral dimensions.
Here 
$s$ is an intensity parameter of the periodic optical potential,
$E_\mathrm{R} = \frac{\hbar^2}{2m}\left(\frac{\pi}{d}\right)^2$ is a recoil energy,
$m$ is the mass of a $^{87}$Rb atom, $\nu_r$ and $\nu_z$ are the frequencies of the
harmonic potentials, and $d$ is the period of the potential.
We employ $d=0.395 \, \mu\mathrm{m}$ which is one-half of
the typical~\cite{scott}
wavelength $\lambda$ for the laser beams, $\nu_r = 100 \, \mathrm{s}^{-1}$
and $\nu_z = 30 \, \mathrm{s}^{-1}$ which are also within the typical ranges 
in experiments.

The condensate may have finite angular momentum
which is essential to treat the systems having vortices.~\cite{martikainen-stoof}
For simplicity, we assume the condensate rotational symmetry
\begin{equation}
    \phi(r, z, \theta) = \phi(r, z)e^{iw\theta},
\label{eq:winding}
\end{equation}
where $w = 0$ for vortexfree system and $w=1$ for systems with a vortex.
The wavefunction $\phi(r, z)$ is a real function.
If this function is complex and has vorticity in $rz$ plane,
the condensate will have parallel vortex rings~\cite{crasovan}
which is also an interesting subject.

The BEC in an optical lattice with the lower value of 
the intensity $s$
is well described by the Gross-Pitaevskii equation~\cite{pethick-smith}
\begin{equation}
    \left( - C \nabla^2 + V(r, z) + g|\phi(\bm{r})|^2 \right)\phi(\bm{r})
    = \mu\phi(\bm{r})
\label{eq:gp}
\end{equation}
where $\phi(\bm{r})$ is the wavefunction of the condensate,
$\mu$ is the chemical potential,
$g = \frac{4\pi\hbar^2a}{m}$ is an interaction parameter,
and $C = \hbar^2/(2m)$ is a constant.
We denote the components of $\bm{r}$ to be $(r, z, \theta)$.
We treat a condensate with $5\times 10^4$ atoms of $^{87}$Rb.
The scattering length $a = 5.4 \, \mathrm{nm}$ 
is employed for the $^{87}$Rb atoms.
The period of the optical lattice is $d=0.395 \, \mu\mathrm{m}$ and
the condensate occupies about 80 sites.
The centered lattice site ($-d/2<z<d/2$) is occupied by about 1290 (1110) atoms
when $s=0$ ($s=10$). 

The condensate supports excitation spectra within the Bogoliubov framework.
The Bogoliubov equations~\cite{pethick-smith}
\begin{align}
    \left(- C\nabla^2 + V(r, z) + 2g|\phi|^2 - \mu\right)u_q
        - g\phi^2v_q \!\!&=&\!\! \varepsilon_q u_q,
\label{eq:bdg1}
\\
    \left(- C \nabla^2 + V(r, z) + 2g|\phi|^2 - \mu\right)v_q
        - g\phi^{\ast 2}u_q \!\!&=&\!\! -\varepsilon_q v_q
\label{eq:bdg2}
\end{align}
yield the excitation energies $\varepsilon_q$
and the corresponding wavefunctions, $u_q$ and $v_q$.
The symmetry of system enables us to rewrite the wavefunctions
\begin{align}
    u_q(\bm{r}) &= u_q(r, z)e^{i(w + q_{\theta})\theta},
\\
    v_q(\bm{r}) &= u_q(r, z)e^{i(w - q_{\theta})\theta},
\end{align}
where $u_q(r, z)$ and $v_q(r, z)$ are real functions
and $q_{\theta}$ is an integer representing the angular momentum.
The excitations with angular momentum $q_{\theta}= -1$ is
related to the oscillation of vortex core and are 
called Kelvin modes.

\section{Vortex Oscillations}

\begin{figure}
\begin{center}
\hspace{-1cm}\includegraphics[width=8.5cm,clip]{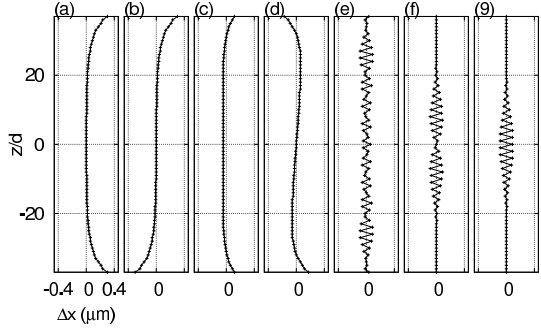}
\vspace{-1cm}

\end{center}
\caption{\label{fig:fluc1}
Displacement of the vortex core at the centers ($z = jd$, $j$ is an integer)
of the optical lattice sites at $s=5$.
(a) - (d) $c = 0 \mathrm{\  to\ } 3$, (e) $c = 64$, (f) $c = 51$, and (g) $c = 36$.
}
\end{figure}

We consider the small perturbation around the stationary state
\begin{equation}
\phi^{\prime}(\bm{r}, t) = 
\phi(\bm{r}) + C_1\left(
    u_q(\bm{r})e^{-i\varepsilon_qt/\hbar} - v_q^{\ast}(\bm{r})e^{i\varepsilon_qt/\hbar}
\right)
\label{eq:perturbation}
\end{equation}
where $C_1$ is a constant.
At $t = 0$, the perturbation by the Kelvin modes ($q_{\theta}= -1$) is
\begin{equation}
    \phi^{\prime}(\bm{r},0) = \phi(r, z)e^{i\theta} + 
    C_1 \left( u_q(r, z) - v_q(r, z)e^{-2i\theta} \right)
    \label{eq:at:t0}
\end{equation}
and the corresponding modified density $n^{\prime} \equiv |\phi^{\prime}|^2$ at $t=0$ is
\begin{multline}
    n^{\prime}(r, z, \theta)
         \simeq  n(r, z) + 2 C_1 \phi(r, z)
\\
  \times \left\{ u_q(r, z) \cos \theta  - v_q(r, z) \cos 3\theta \right\}
    \label{eq:perturbation_dns}
\end{multline}
where $n \equiv |\phi|^2$. 
The principal direction of perturbation of the density $n^{\prime}(r, z, \theta)$ above
is along the $x$-axis ($\theta = 0 \mathrm{ and } \pi$).
We write the displacement of the vortex core in the $xz$ plane $\Delta x(z)$,
which is defined at bottoms of the lattice sites $z = jd$ ($j$ is an integer).
Figures \ref{fig:fluc1}(a) - (g) show several patterns of the
displacement $\Delta x(z)$ of the vortex core.

To balance the original density and the perturbation, we employ a value
\begin{equation}
    C_1 =
    0.1 \frac{ \max_{\bm r}\{ n(\bm{r}) \} }{ 
        \max_{q, \bm{r}}\left[
            | \phi(\bm{r}) \left\{ u_q(\bm{r}) - v_q(\bm{r}) \right\} |
        \right]
    }.
\label{eq:c_prime}
\end{equation}
in the figures.
Here $q$ covers the Kelvin modes $(q_\theta=-1)$.


\begin{figure}
\begin{center}
(a)
\includegraphics[width=8cm,clip]{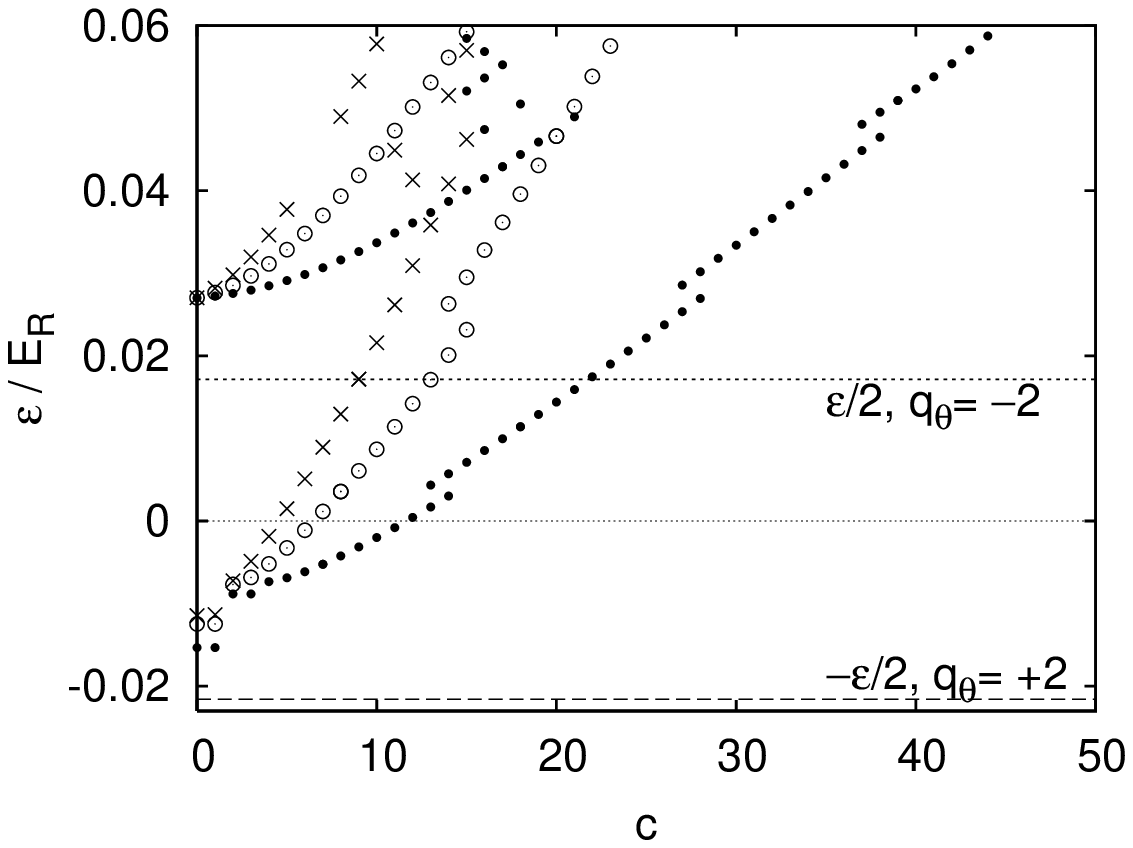}

(b)
\includegraphics[width=8cm,clip]{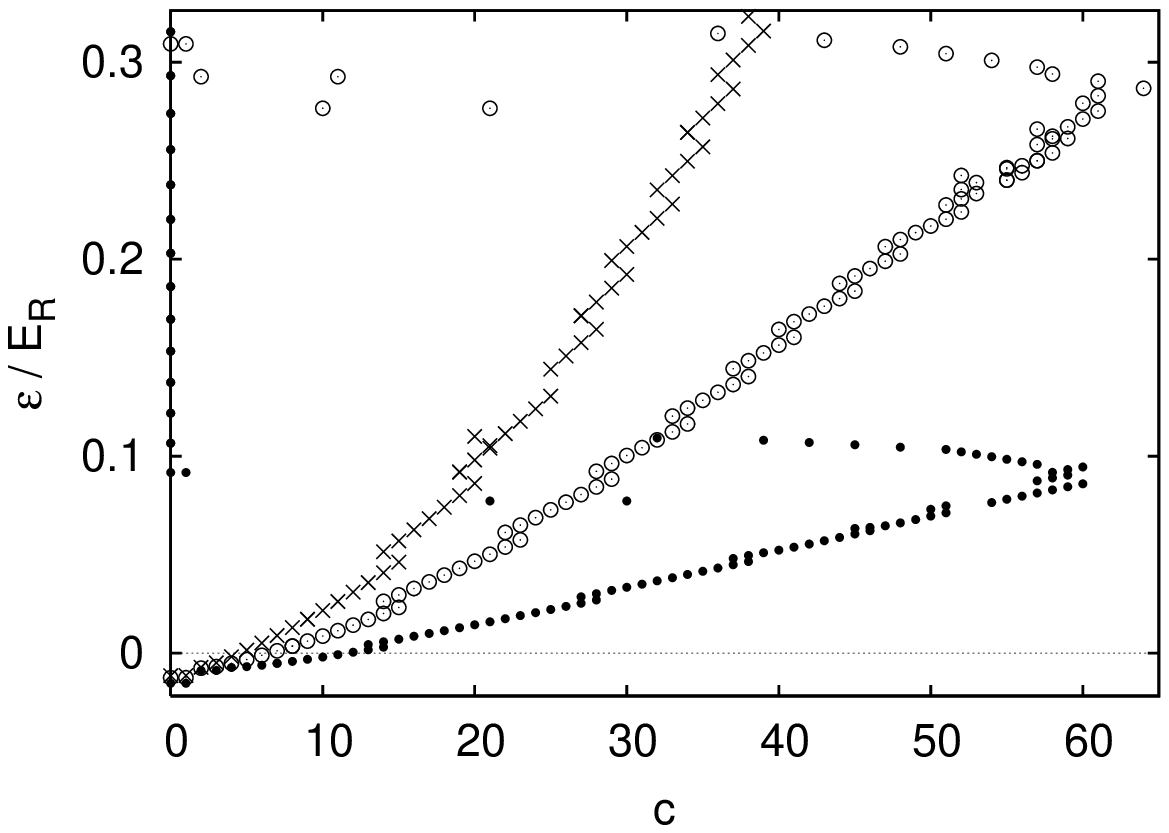}

\end{center}
\caption{\label{fig:core_and_quad}
(a)(b) Displacement count $c$ of the vortex core for $s=0$ ($\times$), 
$s=5$ ($\circ$), and $s=10$ ($\bullet$).
(a) The upper horizontal dotted line indicates
half of the energy of counter-rotating quadrupole mode
[see Fig.~\ref{fig:vtx_quadrupole_both}(f)].
The dashed horizontal line shows the energy 
$- \varepsilon / 2$ where $\varepsilon$ here is the energy of the
lowest co-rotating quadrupole mode [Fig.~\ref{fig:vtx_quadrupole_both}(a)].
(b) The same plot with wider ranges of $c$ and $\varepsilon$.
The modes whose wavefunction $u_q$ has node along the radial direction
are omitted.
}
\end{figure}

On unconfined periodic system, a wavenumber, {e.g.} $k$, 
denotes the phase of the wavefunction through a factor $\mathrm{exp}(ikz)$.
But explicitly, this definition of $k$ cannot be used in confined systems.
The exception is the condensate interacting with moving optical lattices.
The velocity of the optical lattice and the wavenumber are closely
related, and the wavenumber remains good approximation to
denote a excitation.~\cite{modugno,pitaevskii-stringari,kirstine}
Because the moving optical lattice is not interest in this article,
we employ another index similar to the wavenumber, but
better related to an observable property of the excitation.

We define a number $c$ that counts the change of sign of $\Delta x$.
Because $v_q(r, z) = 0$ at $r=0$,
the sign and the amplitude of $\Delta x$ corresponds
to those of $u_q(r, z)$ at $r=0$.
Figure \ref{fig:core_and_quad}(a) plots displacement count $c$
vs.~excitation energy $\varepsilon$
for system with (bullets, circles) and without (crosses)
the optical lattice potential.
As the optical lattice potential becomes stronger ($s$ increases),
the excitation energy for certain $c$ becomes lower.
It means that vortex oscillations with many nodes are
more likely to be excited at stronger lattice potential.
%

\section{Shortest wavelength modes}

The displacement count $c$ of the Kelvin modes smoothly increase
as shown in Fig.~\ref{fig:core_and_quad}(a).
But this increase breaks down and the $c$ decreases at higher energy,
see Fig.~\ref{fig:core_and_quad}(b).
The largest displacement count is around $c \simeq 60$, which is close to the
number of the occupied sites.

Excitations above the breaking down energy have common features
in their wavefunctions and the corresponding oscillation patterns.
The oscillation ranges whole the condensate [Fig.~\ref{fig:fluc1}(e)]
at the breaking down, but at the higher energies
the $z$ ranges of the oscillations shrink. 
Accompanying this shrink, the displacement count $c$
decrease to, for example, 36 [Fig.~\ref{fig:fluc1}(f) - (g)].

These plots show that the displacement $\Delta x$ in this decreasing aspect
has opposite sign between the neighboring sites.
It means that 
the wavefunctions $u$ at the vortex core ($r=0$)
have opposite phase between the neighboring sites.
Therefore, the wavelength of these excitations are $2d$.
This is the shortest wavelength possible 
within the range of excitation energy $\varepsilon \ll s E_\mathrm{R}$ we consider.
We call these excitations the shortest wavelength excitations hereafter.

\begin{figure}
\begin{center}
\includegraphics[width=8cm,clip]{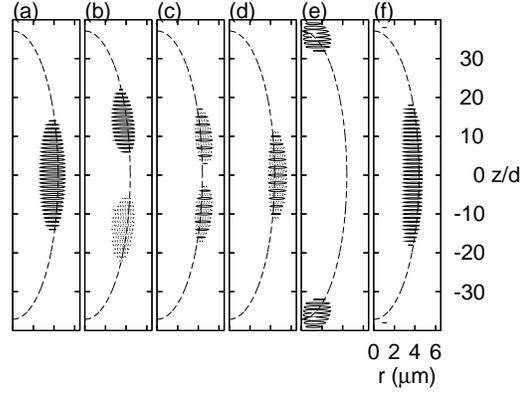}
\vspace{-1cm}

\end{center}
\caption{\label{fig:vtx_quadrupole_both}
Wavefunctions $u(r, z)$ of the quadrupole ($q_{\theta} = +2, -2$) modes
at intensity $s=5$.
Solid (dotted) lines show contours at $u(r, z) > 0.01$ ($u(r, z) < -0.01$).
Dashed lines are Thomas-Fermi profile of the vortexfree condensate
without the optical lattice.
(a) - (b) First two excitations within $q_{\theta} = +2$.
(c) - (d) Two excitations with highest energy within the first band
($q_{\theta} = +2$, radial node free).
Note that each of these two have opposite sign (\textit{i.e.} opposite phase)
of $u$ between the neighboring sites.
(e) The first excitation within $q_{\theta} = -2$.
(f) The first centered excitation of $q_{\theta} = -2$.
Vortexfree [$w=0$ in eq.~(\ref{eq:winding})] condensate 
supports quadruple ($q_{\theta} = \pm 2$) modes similar to these (a) - (d).
}
\end{figure}

We pointed out the shortest wavelength excitations within the Kelvin modes.
In addition to these, excitations of this kind commonly exist
in the confined condensates interacting with an optical lattice.
The condensate with and without the vortex supports
shortest wavelength excitations within the breathing ($q_{\theta}=0$),
dipole ($q_{\theta}=\pm 1$), quadrupole ($q_{\theta}=\pm 2$) modes,
and modes with several higher angular momenta.
Figures ~\ref{fig:vtx_quadrupole_both}(c) and \ref{fig:vtx_quadrupole_both}(d)
are those within the
quadrupole modes of $q_{\theta}=+2$.
One-dimensional system with 1D optical lattice and
harmonic confinement potential also has corresponding excitations.
The set of these excitations are mentioned as ``branch B" in ref.~\citen{iso-lattice}.

\begin{figure}
\begin{center}
(a)
\includegraphics[width=8cm,clip]{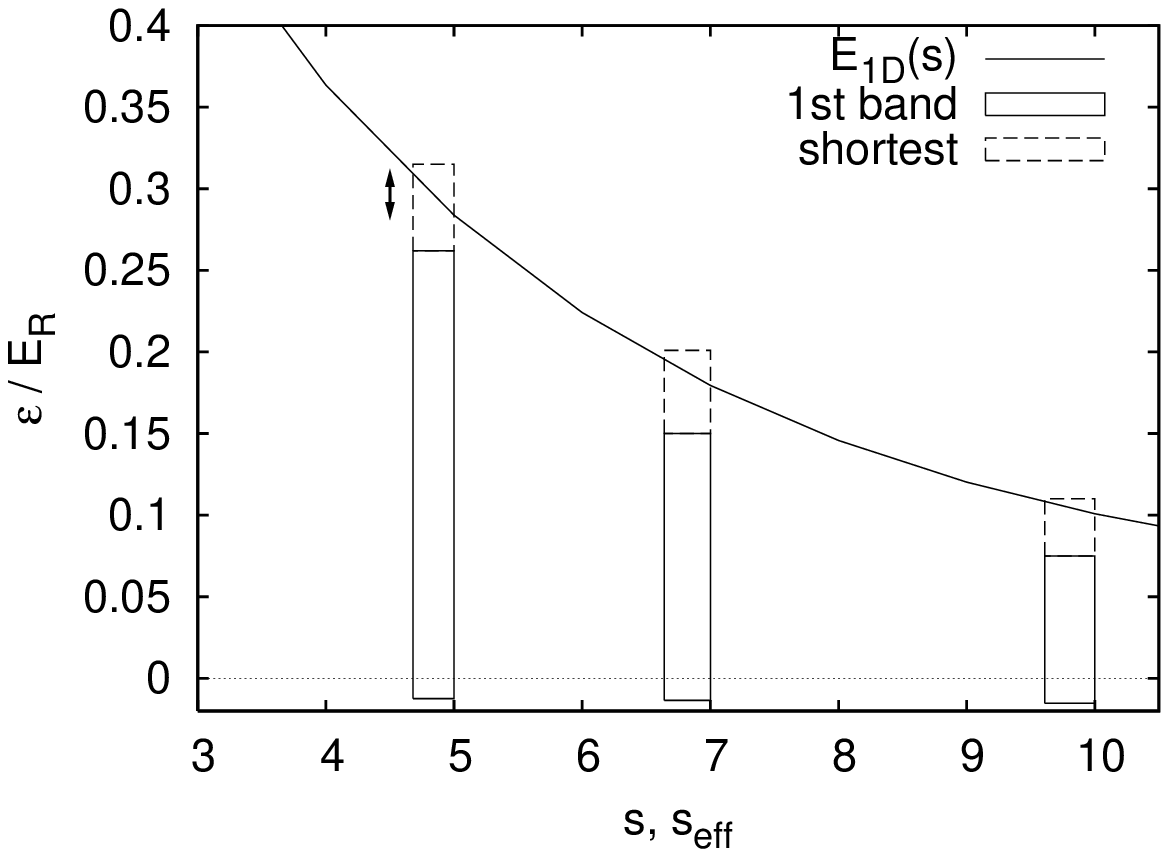}

(b)
\includegraphics[width=8cm,clip]{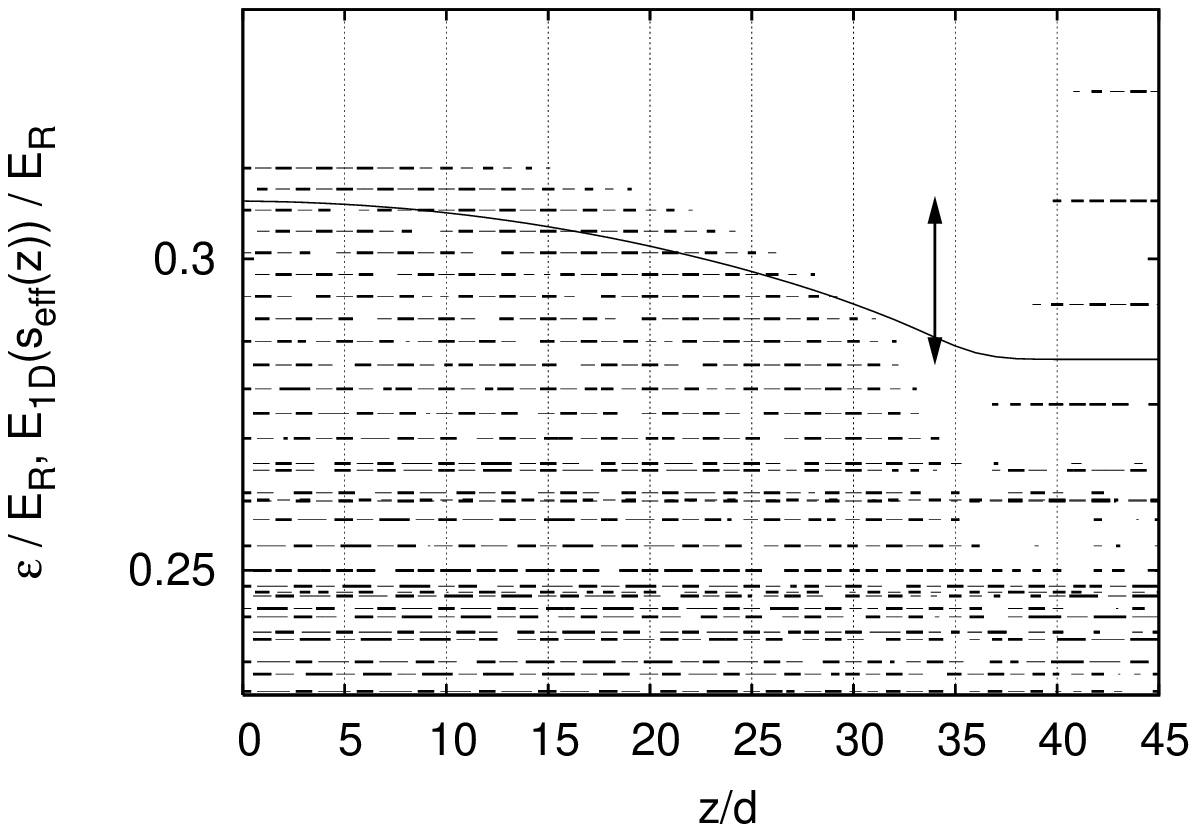}

\end{center}
\caption{\label{fig:firstband}
(a) Boxes shows ranges of energy (vertical) of the Kelvin modes
and $s_\mathrm{eff}$ (horizontal)
for systems with lattice intensities $s = 5, 7, \mathrm{and}\  10$.
Solid curve is energy $E_\mathrm{1D}(s)$ of upper end 
of the first band for the noninteracting 1D periodic system.
(b) Energies $E_\mathrm{1D}(s_{\mathrm{eff}}(z))$ corresponding to
the local value of
the effective lattice intensity $s_\mathrm{eff}(z)$ and
spatial distributions of wavefunctions of the Kelvin modes are compared.
Solid curve is $E_\mathrm{1D}(s_{\mathrm{eff}}(z))$.
Sets of the horizontal short lines are the wavefunctions
$u_q(r, z)$ at $r = 0$ at its energy $\varepsilon_q$.
Bold (thin) lines shows $u_q(0,z)$ is positive (negative).
Axial ($z$-axis) distributions of the Kelvin modes
are localized near the peak of $E_\mathrm{1D}( s_{\mathrm{eff}}(z) )$.
Both of the arrows in (a) and (b) shows
the same range of $E_\mathrm{1D}(s_{\mathrm{eff}}(z))$ for $s = 5$.
}
\end{figure}

\subsection{Energy of the shortest wavelength modes}\label{sec:ene}

In an unconfined 1-dimensional periodic system with period $d$,
the largest wavenumber is $\pi / d$,
and the corresponding phase difference between the neighboring sites is $\pi$
at the upper end of the so-called first band.
The shortest wavelength modes in our confined systems
also have the phase differences $\pi$.
And there is always an energy gap above the shortest wavelength excitations
if we ignore the modes with radial nodes.
So it is reasonable to think that these excitations also
belong to the upper end of the first band.

Figure \ref{fig:firstband}(a) compares
ranges of the energy of the shortest wavelength excitations
and the upper end $E_\mathrm{1D}(s)$ of the first band of
an ideal (noninteracting, unconfined)
1-dimensional periodic system.~\cite{pitaevskii-stringari,kirstine,iso-lattice}
It is seen that the energy ranges of the shortest wavelength excitations overlap
with $E_\mathrm{1D}(s)$.
The energy ranges of the
shortest wavelength excitations of another angular momenta $q_\theta = 0, 1, \pm2$
also overlap $E_\mathrm{1D}(s)$ despite the change in the angular momenta.
So the difference of angular momentum is not essential in the comparison in 
Fig.~\ref{fig:firstband}(a).

Range of $E_\mathrm{1D}$ is $0 < E_\mathrm{1D} < E_\mathrm{R}$ 
(numerically calculated from eq.~(12) in ref.~\citen{iso-lattice}).
So the ranges of the shortest wavelength modes for various $s$ will also
be within $0 < E_\mathrm{1D} < E_\mathrm{R}$.

\subsection{Origin of the spatial dependence}\label{sec:spatial}

While other modes are elongated whole the condensate,
he shortest wavelength excitations are localized 
near the center ($z=0$) of the system.
This feature is clearer for modes with the highest energies,
\textit{e.g.} Figs.~\ref{fig:fluc1}(g) and \ref{fig:vtx_quadrupole_both}(d).

Within the GP equation and the Bogoliubov equations, 
the depth of the optical lattice $s$ is changed due to the meanfield energy $g|\phi|^2$ in
eqs.~(\ref{eq:gp}), (\ref{eq:bdg1}), and (\ref{eq:bdg2}).
The effective depth becomes smallest near the center ($z=0$) of the system.
In 1D periodic system, the smaller $s$ becomes,
the higher the upper edge $E_\mathrm{1D}(s)$ of the first band becomes.
This agrees with the localization of the highest shortest wavelength mode
around $z=0$ at which the effective $s$ becomes the smallest.

To make this discussion clear, we define effective amplitude 
\begin{equation}
    s_\mathrm{eff}(z) = s - \max_{r}\left( g|\phi(r, z)|^2 \right) / E_\mathrm{R}
\end{equation}
at bottoms of the lattice sites $z = jd$ ($j$ is an integer).
Within the `$\max$' above, the radius $r$ is chosen to have the maximum.
The function $s_\mathrm{eff}$(z) is smallest near the center ($z=0$) of the system.

Spatial (along $z$-axis) dependence of the effective depth $s_\mathrm{eff}(z)$
and the energy may be related by a function $E_\mathrm{1D}(s_\mathrm{eff}(z))$.
This function and the energy and spatial localization of the Kelvin modes
are compared in  Fig.~\ref{fig:firstband}(b).
The excitations at the left hand side
above $\varepsilon > 0.27\, E_R$ are the wavefunctions of the
shortest wavelength modes.~\cite{explain}
The correspondence between the
$s_\mathrm{eff}$ and the axial ($z$-) range of the wavefunctions
is clear.

\section{How to observe the shortest wavelength modes}

\begin{figure}
\begin{center}
\includegraphics[width=8cm,clip]{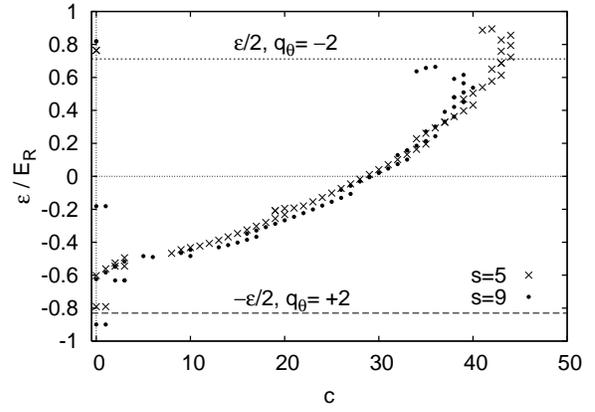}
\end{center}
\caption{\label{fig:nodes:4000hz}
Displacement count $c$ vs.~energy of the Kelvin modes
for $s=5$ ($\times$) and $s=9$ ($\bullet$).
The trapping potential parameters are
$\nu_r = 4000\, \mathrm{Hz}$, $\nu_z = 300\, \mathrm{Hz}$.
The horizontal lines are the same as Fig.~\ref{fig:core_and_quad}(a).
}
\end{figure}

On trapped BEC of atomic gases,
the quadrupole modes may be excited experimentally~\cite{kelvon} by deforming
the confining potential.
And both the experiment~\cite{kelvon} and
the following theoretical study~\cite{kelvon:theory}
insist that Kelvin modes ($q_\theta = -1$) are
excited by the decay from
the counter-rotating quadrupole mode ($q_\theta = -2$).
For the condensates interacting with the periodic potentials,
a resonance condition between the quadrupole mode and the Kelvin modes
is given in \S~V of ref.~\citen{martikainen-stoof}.
The criteria given in the references above may be simplified to this:
The excitation energy $\varepsilon_{-1}$
(or frequency $\varepsilon_{-1}/\hbar$) of the excited Kelvin modes satisfy 
\begin{equation}
     \varepsilon_{-1} = \varepsilon_{-2}/2,
    \label{eq:halfenergy}
\end{equation}
where $\varepsilon_{-2}$ is the energy of the
counter-rotating quadrupole mode at $q_{\theta} = -2$.
If one of the shortest wavelength modes satisfies eq.~(\ref{eq:halfenergy}),
the mode will be excited as an oscillation of the vortex core.
The oscillation will be experimentally observable.

The corresponding excitations in this paper is
the lowest centered counter-rotating quadrupole mode $q_{\theta} = -2$
in Fig.~\ref{fig:vtx_quadrupole_both}(f)
and the Kelvin mode whose energy satisfies eq.~(\ref{eq:halfenergy}).
The excitation energy $\varepsilon_{-2}$
varies only little (stays between $0.0343 E_\mathrm{R} \text{ to } 0.0351 E_\mathrm{R}$)
for $s = 0 \text{ to } 10$.
And half of the energy $\varepsilon_{-2}/2$ is indicated in Fig.~\ref{fig:core_and_quad}(a).
The figure shows that 
Kelvin modes with $c = 13$ ($c = 22$) at $s = 5$ ($s = 22$) are 
on the dotted line and satisfying the criterion.

Then how can we make the shortest wavelength modes satisfy eq.~(\ref{eq:halfenergy})?
The right hand side, the energy of the centered counter-rotating quadrupole modes is 
$\varepsilon_{-2} \simeq 1.2 \, h \nu_{r}$ where $h$ is the Plank constant.
The left hand side, the energy of the shortest wavelength modes,
ranges $0 < \varepsilon < E_\mathrm{R}$ from the discussion in \S~\ref{sec:ene}.
As $E_\mathrm{R}$ is a function of period $d$ of the optical lattice
and $\varepsilon_{-2}$ is approximately a function of radial confinement potential
parameter $\nu_{r}$, these two may be varied independently.

We increased $\nu_r$ while keeping $d$.
At $\nu_r = 4000 \, \mathrm{Hz}$, the shortest wavelength mode satisfies eq.~(\ref{eq:halfenergy}).
Here axial trap frequency $\nu_z$ particle number $N$
are changed to $\nu_z = 300 \, \mathrm{Hz}$ and $N = 10^5$.
Figure \ref{fig:nodes:4000hz} presents spectra of the Kelvin modes
and the energy $\varepsilon_{-2}/2$.
In this tight radial trapping potential, the shortest wavelength modes
satisfies eq.~(\ref{eq:halfenergy}) at $s > 5$ and
localized vortex oscillation of period $2d$ will be observed
by exciting the counter-rotating quadrupole mode.

In writing eq.~(\ref{eq:halfenergy}), we neglected the width of the resonance 
whose evaluation~\cite{kelvon:theory} requires huge computational resource.
Explicitly speaking, the condition is satisfied at only discrete values of $s$.
The allowed width of $s$ is a function of various parameters, including the temperature.
This discreteness can be avoided by smoothly varying $s$.

\section{Discussion}

In BEC of atom gases in 1D optical lattice,
existence of the spatially localized modes was reported in ref.~\citen{iso-lattice}.
But the nature of the modes was not clarified there.
In this paper, we find that the modes have common wavelength $2d$ and 
commonly exists at various angular momenta,
and exists both at the vortexfree condensate and the condensate with a vortex.

Among the modes, we report on the Kelvin modes and the
corresponding vortex oscillations in systems with vortex.
Provided that
the vortex oscillation was observed~\cite{kelvon} in systems without the optical lattice,
the same observation will be possible with optical lattice.
By choosing the trapping potentials and the intensity of the optical lattice,
we can make the vortex oscillation modes of wavelength $2d$ satisfy
the resonance condition eq.~(\ref{eq:halfenergy}).
Then the corresponding oscillation, whose directions
are the opposite between the neighboring sites, will be observed.


I thank S. K. Yip and T. Mizushima for useful discussions.


%
%
%
%



\begin{thebibliography}{99}
\bibitem{explain}
The excitations at the right hand side in Fig.~\ref{fig:firstband}(b)
exist outside the condensate
and do not contribute to the vortex oscillations.
\end{thebibliography}
\end{document}